\documentclass[%
 aip,
 jmp,%
 amsmath,amssymb,
 reprint,%
]{revtex4-1}

\usepackage{graphicx}
\usepackage{dcolumn}
\usepackage{bm}
\usepackage{epstopdf}
\usepackage{amssymb}
\usepackage{color}
\usepackage[colorlinks=true, letterpaper=true, pdfstartview=FitV, linkcolor=blue, citecolor=blue, urlcolor=blue]{hyperref}
\usepackage{lineno}

\begin{document}

\preprint{AIP/123-QED}

\title[]{Phonon and electron transport in Janus monolayers based on InSe}

\author{Wenhui Wan}
\author{Shan Zhao}
\author{Yanfeng Ge}
\author{Yong Liu}
\email{ycliu@ysu.edu.cn or yongliu@ysu.edu.cn}
\affiliation{State Key Laboratory of Metastable Materials Science and Technology $\&$
Key Laboratory for Microstructural Material Physics of Hebei Province,
School of Science, Yanshan University, Qinhuangdao, 066004, P.R. China}

\date{\today}

\begin{abstract}
  We systematically investigated the phonon and electron transport properties of monolayer InSe and its Janus derivatives including monolayer In$_{2}$SSe and In$_{2}$SeTe by first-principles calculations. The breaking of mirror symmetry produces a distinguishable $A_{1}$ peak in the Raman spectra of monolayer In$_{2}$SSe and In$_{2}$SeTe. The room-temperature thermal conductivity ($\kappa$) of monolayer InSe, In$_{2}$SSe and In$_{2}$SeTe is 44.6, 46.9, and 29.9 W/(m，K), respectively. There is a competition effect between atomic mass, phonon group velocity and phonon lifetime.
  The $\kappa$ can be further effectively modulated by sample size for the purpose of thermoelectric applications. Meanwhile, monolayer In$_{2}$SeTe exhibits a direct band and higher electron mobility than that of monolayer InSe, due to the smaller electron effective mass caused by tensile strain on the Se side. These results indicate that 2D Janus group-III chalcogenides can provide a platform to design the new electronic, optoelectronic and thermoelectric devices.
\end{abstract}

\maketitle

Indium Selenide (InSe), as an experimentally accessible
layered group-III metal chalcogenide, has been receiving much attention.~\cite{nano7110372}
The synthesized few-layer InSe exhibits small electron effective mass ($\sim$ 0.14 $m_{0}$) and room-temperature electron mobility ($\mu$) higher than $10^{3}$ cm$^{2}$/(V$\cdot$s).~\cite{bandurin2016high,InSehigh2}
Nano-devices based on two dimensional (2D) InSe had good ambient stability.~\cite{stable2}
In addition, 2D InSe holds promise for the application of bendable photodetectors with broadband response,~\cite{tamalampudi2014high} 2D
ferromagnets,~\cite{cao2015tunable} excitonic dynamics,~\cite{dey2014mechanism} magneto-optical effects~\cite{magnetooptical}
and topological insulator.~\cite{Zhou2018}

Recently, a new Janus-type monolayer MoSSe has been synthesized through the replacement of the S atoms at one side of monolayer MoS$_{2}$ by Se atoms.~\cite{Lu2017,janus1}
Both band gap and thermal conductivity $\kappa$ of monolayer MoSSe are between that of monolayer MoS$_{2}$ and MoSe$_{2}$.~\cite{C8CP00350E}
As far as Janus structures based on 2D InSe are concerned,
Kandemir \textit{et al.} found that the band structure of the monolayer In$_{2}$SSe was formed by the superposition of the strained band structures of the binary analog single layers.~\cite{InSeS}
Guo \textit{et al.} predicted that the 2D Janus group-III chalcogenides had enhanced piezoelectric coefficients compared to that of perfect ones.~\cite{piezoelectric}
The out-of-plane symmetry breaking in 2D Janus structures can also induce unusual properties such as enhanced Rashba effect,~\cite{rashba} catalytic activity for hydrogen evolution,~\cite{Catalytic} the valley polarization upon magnetic doping,~\cite{Valley} excitons with long lifetime~\cite{C8NR04568B} and so on.
Besides the electronic properties, thermal properties become a crucial issue for the device's performance, as on-going miniaturization of electronic devices.
For example, a high and low $\kappa$ is desirable for the efficient heat dissipation in integrated electronic devices and high conversion efficiency of thermoelectrics, respectively.~\cite{silinano1}
Nissimagoudar \emph{et al.} predicted that the $\kappa$ of monolayer InSe can be effectively reduced by boundary scattering.~\cite{InSe}
Pandey \emph{et al.} calculated the $\kappa$ of monolayer GaS, GaSe, and InSe, and found that the increase of mass gave both decreasing
acoustic phonon velocities and increasing scattering of heat-carrying modes.~\cite{InSe2}
The thermal properties of 2D materials can be tuned by doping or alloying,~\cite{C6NR04651G} strain,~\cite{DENG201814} chemical functionalization~\cite{tune2017} and so on, expanding its application prospect.
Therefore, Janus structures can not only expand the family of 2D materials but also offer another way to modulate the thermal properties of 2D materials.
Though much efforts have been devoted to the physical properties of 2D Janus group-III chalcogenides, the mechanism of its phonon and electron transport are still not well understood.

In this work, we studied the structural, optical and transport properties of monolayer InSe, In$_{2}$SSe and In$_{2}$SeTe by first-principles calculations. Because of the breaking of mirror symmetry, Raman spectra of Janus structures exhibit a special $A_{1}$ peak compared to that of monolayer InSe. The $\kappa$ of monolayer In$_{2}$SSe and In$_{2}$SeTe is larger and smaller than that of monolayer InSe. The mechanism of heat transport are carefully analyzed. At last, we calculated the electron mobility $\mu$ of monolayer InSe and its Janus structures. The $\mu$ of monolayer In$_{2}$SeTe is superior to that of monolayer InSe.

\begin{table*}[tb]
\caption{\label{table1}
The lattice constant ($a$), cation-cation bond length ($d_{cc}$), anion-cation bonding length ($d_{ac}$), effective thickness of monolayer ($l$), band gap ($E_{g}$), the frequency and irreducible representations of Raman peak of monolayer InSe, In$_{2}$SSe and In$_{2}$SeTe. The previous results~\cite{Zhou2018} of monolayer InSe
are displayed in bracket for a comparison.}
\begin{ruledtabular}
\begin{tabular}{ccccccccc}
      &$a$(\AA) &$d_{cc}$(\AA) &$d_{ac}$(\AA) &$l$(\AA) & $E_{g}(eV)$ &\multicolumn{3}{c}{$\omega_{Raman}$(cm$^{-1}$)} \\
\hline
InSe         &4.093	 &2.817	 &2.690 &5.386 &2.147 &109 ($A_{1}^{'}$)[110]~\cite{Zhou2018} &178 ($E^{'}$)[178]~\cite{Zhou2018} &223 ($A_{1}^{'}$)[225]~\cite{Zhou2018} \\
In$_{2}$SSe 	&4.015		 &2.819			 &2.671/2.585	&5.289 &2.297 & 124 ($A_{1}$)     & 214 ($A_{1}$) & 257 ($A_{1}$) \\
In$_{2}$SeTe	&4.244		 &2.815			 &2.729/2.856	&5.483 &1.967 &  96 ($A_{1}$)      & 167 ($A_{1}$) & 211 ($A_{1}$) \\
\end{tabular}
\end{ruledtabular}
\end{table*}

\begin{figure}[tbp!]
\centerline{\includegraphics[width=0.45\textwidth]{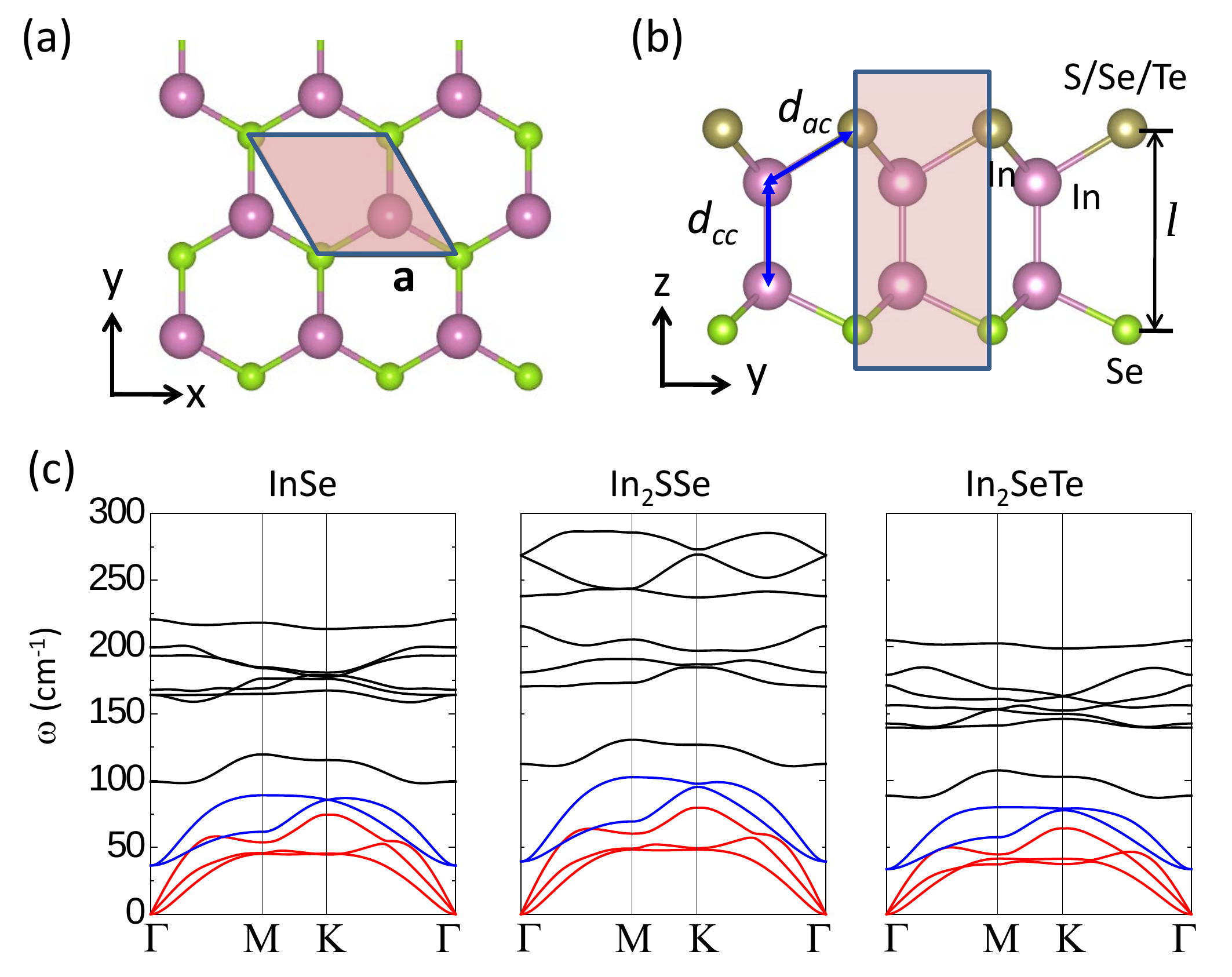}}
\caption{(a) A top view and (b) a side view of crystal structure of monolayer InSe, In$_{2}$SSe and In$_{2}$SeTe, with the shaded area showing the unit cell. The $d_{cc}$ and $d_{ac}$ is the cation-cation and anion-cation bonding length, respectively. (c) The phonon dispersion of monolayer InSe, In$_{2}$SSe and In$_{2}$SeTe.}
\label{wh1}
\end{figure}

Based on the phonon Boltzmann transport function, lattice thermal conductivity $\kappa$ is estimated by~\cite{li2014shengbte}
\begin{eqnarray} \label{kappa}
\kappa_{\alpha\beta}= \frac{1}{N\Omega}\sum\limits_{
\mathbf{q},s} {C_{\mathbf{q},s}v_{\mathbf{q},s}^{\alpha}v_{\mathbf{q},s}^{\beta}\tau_{\mathbf{q},s}},
\end{eqnarray}
where $\alpha$ and $\beta$ are Cartesian indices. $N$ and $\Omega$ is the number of $\mathbf{q}$ point and volume of the unit cell, respectively. $C_{\mathbf{q},s}$, $v_{\mathbf{q},s}^{\alpha}$ and $\tau_{\mathbf{q},s}$ is the specific capacity, group velocity and lifetime in the single-mode relaxation time approximation (RTA) of the phonon with wavevector $\mathbf{q}$ and branch index $s$, respectively. The phonon lifetime $\tau_{\mathbf{q},s}$ was estimated by combining the anharmonic scattering, isotopic impurities scattering and boundary scattering according to the Matthiessen rule~\cite{li2014shengbte}
\begin{eqnarray} \label{tau}
\frac{1}{\tau_{\mathbf{q},s}}=\frac{1}{\tau_{\mathbf{q},s}^{an}}+\frac{1}{\tau_{\mathbf{q},s}^{iso}}+\frac{1}{\tau_{\mathbf{q},s}^{b}},
\end{eqnarray}
where boundary roughness scattering rate is $1/\tau_{\mathbf{q},s}^{b}=|v_{\mathbf{q},s}|/L$ with $L$ be the sample size.
All the computational details are given in the supplementary material (SM).

Monolayer InSe consists of a quadruple layer in a stacking sequence of Se-In-In-Se and its lattice has $D_{3h}$ crystal symmetry (see Fig.~\ref{wh1}(a) and~\ref{wh1}(b)). Janus monolayer In$_{2}$SSe and In$_{2}$SeTe are built by replacing top-layer Se by S or Te atoms. Janus structures have $C_{3v}$ crystal symmetry without out-of-plane mirror symmetry ($z\rightarrow -z$). The thickness of monolayer was estimated by the distance between the atoms on the outmost layer (see Fig.~\ref{wh1}(c)).
The optimized structural parameters are listed in Table~\ref{table1}. The cation-cation bonding ($d_{cc}$) have a slight change compared to that of monolayer InSe.
The anion-cation bonding length ($d_{ac}$) of In$_{2}$SSe and In$_{2}$SeTe is smaller and larger than that of monolayer InSe, respectively, due to the smaller and larger atomic radius of S and Te atom than that of Se atom. As a result, one side of Janus structures experiences a tensile or compressive strain, compared to its perfect binary analogs.~\cite{InSeS} For example, the bonding length of $d_{In-Se}$ on Se side and $d_{In-Te}$ on Te side of monolayer In$_{2}$SeTe is larger and smaller than that of monolayer InSe and InTe ($d_{In-Te}$ = 2.890 \AA\ in our calculation), respectively (see Table~\ref{table1}).

\begin{figure}[tbp!]
\centerline{\includegraphics[width=0.45\textwidth]{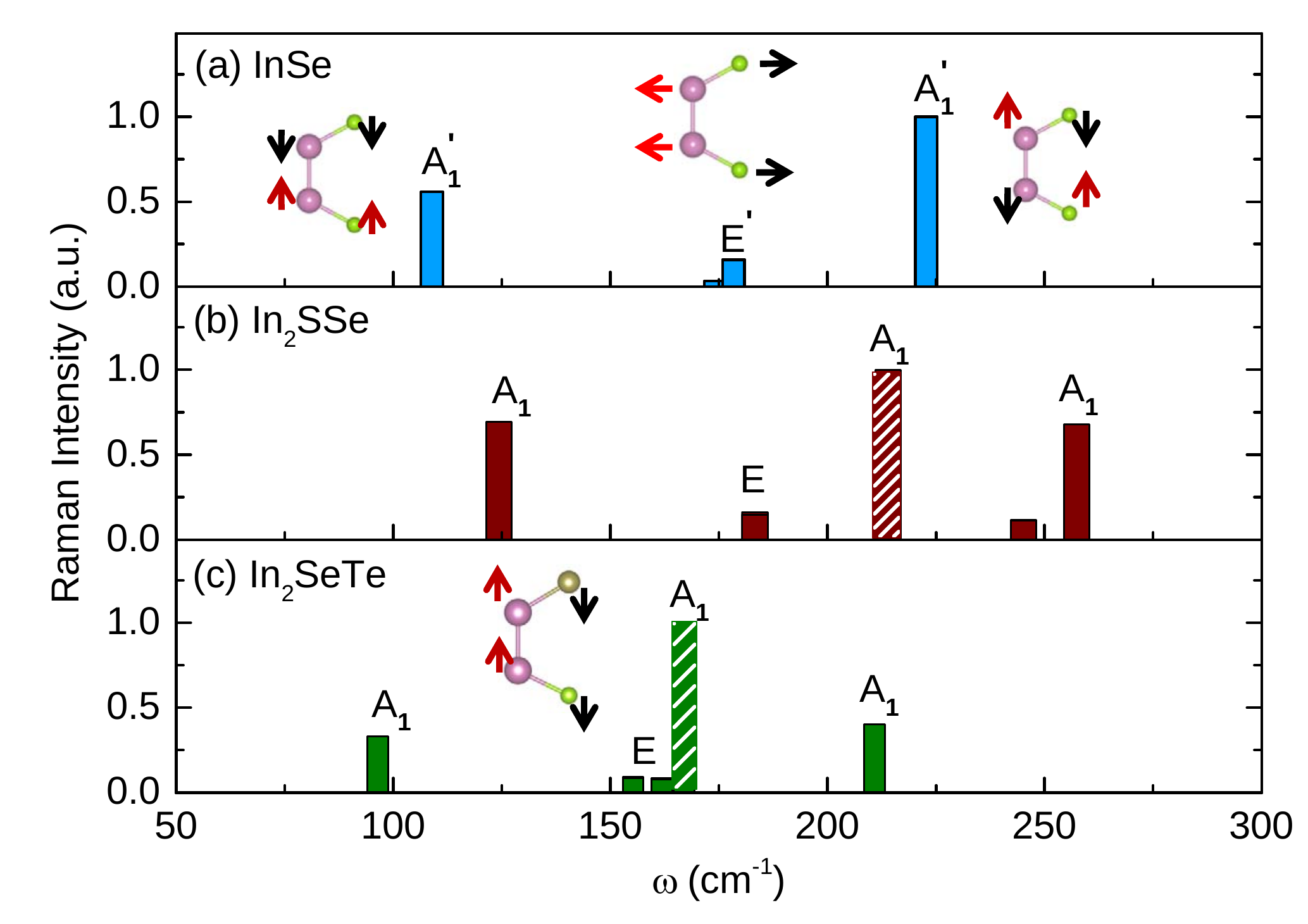}}
\caption{The Ranman spectra of (a) monolayer InSe, (b) In$_{2}$SSe and (c) In$_{2}$SeTe. The eigenvectors of $A_{1}^{'}$ and $E$ mode of monolayer InSe and the $A_{1}$ mode of monolayer In$_{2}$SeTe are displayed.}
\label{wh2}
\end{figure}

Monolayer InSe and its Janus derivatives are semiconductors (see Fig. S2). Figure~\ref{wh1}(c) displays the phonon dispersive relations and indicates their structural stability. The acoustic branches consist of the longitudinal (LA) branch, transverse (TA) branch and flexural branch (ZA) branch. Acoustic branches intersect with optical branches.
The phonon modes at the $\Gamma$ point of monolayer InSe are decomposed into
$\Gamma_{D_{3h}} = 2A^{'}_{1}\oplus 2A^{''}_{2}\oplus 2E^{'}\oplus 2E^{''}$ according to the analysis of group theory. $A^{'}_{1}$, $E^{'}$ and $E^{''}$ modes are Raman active. The calculated Raman spectra exhibits three prominent peaks (see Fig.~\ref{wh2}(a)). Two Raman peaks at frequencies $\omega$ of 109 and 223 cm$^{-1}$ arise from two out-of-plane $A_{1}^{'}$ modes and a smaller peak at $\omega$=178 cm$^{-1}$ originates from in-plane $E^{'}$ mode, consistent with the previous work (see Table~\ref{table1}).~\cite{Zhou2018}

On the other side, the $\Gamma$-point phonon modes of Janus monolayer In$_{2}$SSe and In$_{2}$SeTe are decomposed into $\Gamma_{C_{3v}} = 4A_{1}\oplus 4E$. All the optical modes are Raman active. There are two prominent $A_{1}$ peaks at the low- and high-frequency zone (see Fig.~\ref{wh2}(b) and~\ref{wh2}(c)). Its eigenvectors are similar to that of two $A_{1}^{'}$ peak of monolayer InSe. In addition, a conspicuous $A_{1}$ mode, which is absent in monolayer InSe, appears in the intermediate frequency zone (labeled by slant lines).
The phonon mode with the similar eigen-displacements (see inset of Fig.~\ref{wh2}(c)) in monolayer InSe belongs to $A^{''}_{2}$ representation. The change of polarizability with corresponding normal coordinate at the equilibrium configuration is zero due to the mirror symmetry, so the $A^{''}_{2}$ mode is Raman inactive.
Therefore, the $A_{1}$ raman peak due to the breaking of mirror symmetry can serve as the characteristic signal of 2D In$_{2}$SSe and In$_{2}$SeTe during its synthesis.

\begin{figure}[tbp!]
\centerline{\includegraphics[width=0.5\textwidth]{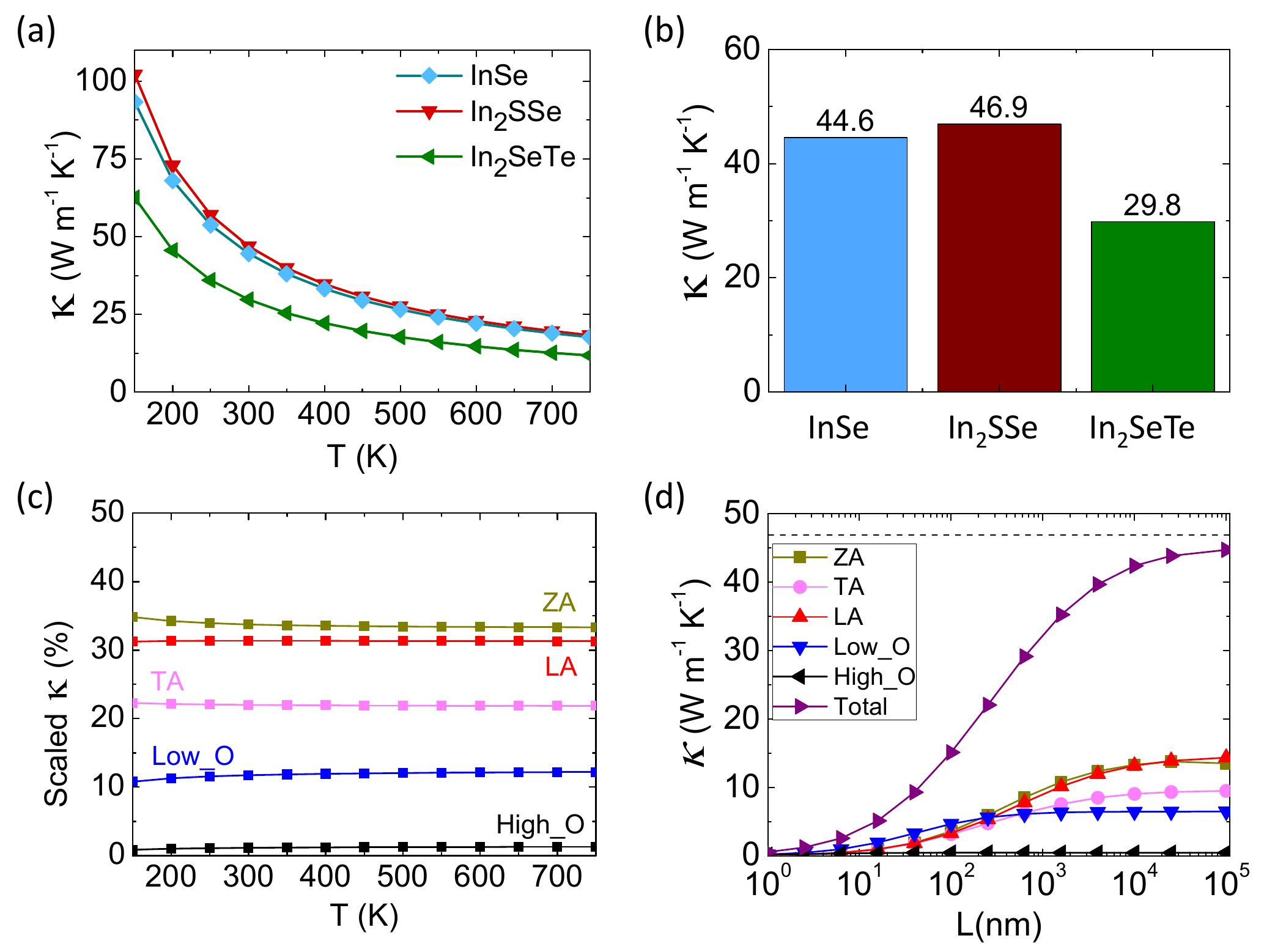}}
\caption{(a) The temperature dependence of the $\kappa$ of monolayer InSe, In$_{2}$SSe and In$_{2}$SeTe. (b) The corresponding room-temperature $\kappa$. (c) The temperature dependence of contribution of $\kappa$ from different branches in monolayer In$_{2}$SSe. (d) The sample size $L$ dependence of the $\kappa$ contributed by different branches in monolayer In$_{2}$SSe.}
\label{wh3}
\end{figure}

The $\kappa$ of monolayer InSe and its Janus derivatives in the temperature range of T = 150$\sim$750 K are shown in Fig.~\ref{wh3}(a). At T = 300 K, the $\kappa$ of monolayer InSe and In$_{2}$SSe is estimated as 44.6 and 46.9 W/(m$\cdot$K), respectively (see Fig.~\ref{wh3}(b)), which is comparable to that of semiconductor GaAs (45 W/(m，K)),~\cite{Gekappa} few-layer MoS$_{2}$ (40$\sim$50 W/(m，K))~\cite{MoS2} and few-layer black phosphorus (10$\sim$20 W/(m，K)).~\cite{Luo2015} The result of monolayer InSe agrees with previous Pandey's work.~\cite{InSe2} In contrast, monolayer In$_{2}$SeTe with heavier atomic mass has a smaller $\kappa$ of 29.8 W/(m$\cdot$K).

Besides the acoustic branches, low-frequency optical ($low$-$o$) branches (labeled by blue in Fig.~\ref{wh1}(e)) also make a non-negligible contribution to the $\kappa$ of monolayer InSe and its Janus derivatives (see Fig.~\ref{wh3}(c)). The large phonon group velocity of $low$-$o$ branches is the main reason. In contrast, only the acoustic phonons are considered in the heat transport of conventional semiconductors. Meanwhile, the heat is mainly carried by phonons with frequencies less than 100 cm$^{-2}$, since the group velocities of higher optical modes are small.

Considering the limited sample size ($L$), the size effects of $\kappa$ and its contribution from different branches are discussed in the range of diffusive thermal transport. The representative result of monolayer In$_{2}$SSe is displayed in Fig.~\ref{wh3}(d). The results of monolayer InSe and In$_{2}$SeTe are similar.
At room temperature, the phonon mean free path (MFP) of monolayer In$_{2}$SSe is about $10^{5}$ nm, which is mainly determined by the MFP of LA and ZA phonons. The $\kappa$ can be decreased by 90\% as the $L$ decrease down to 10 nm. That indicates that nanostructuring might be an effective method to reduce the $\kappa$ of 2D group-III chalcogenide for the thermoelectric applications.

\begin{figure}[tbp!]
\centerline{\includegraphics[width=0.5\textwidth]{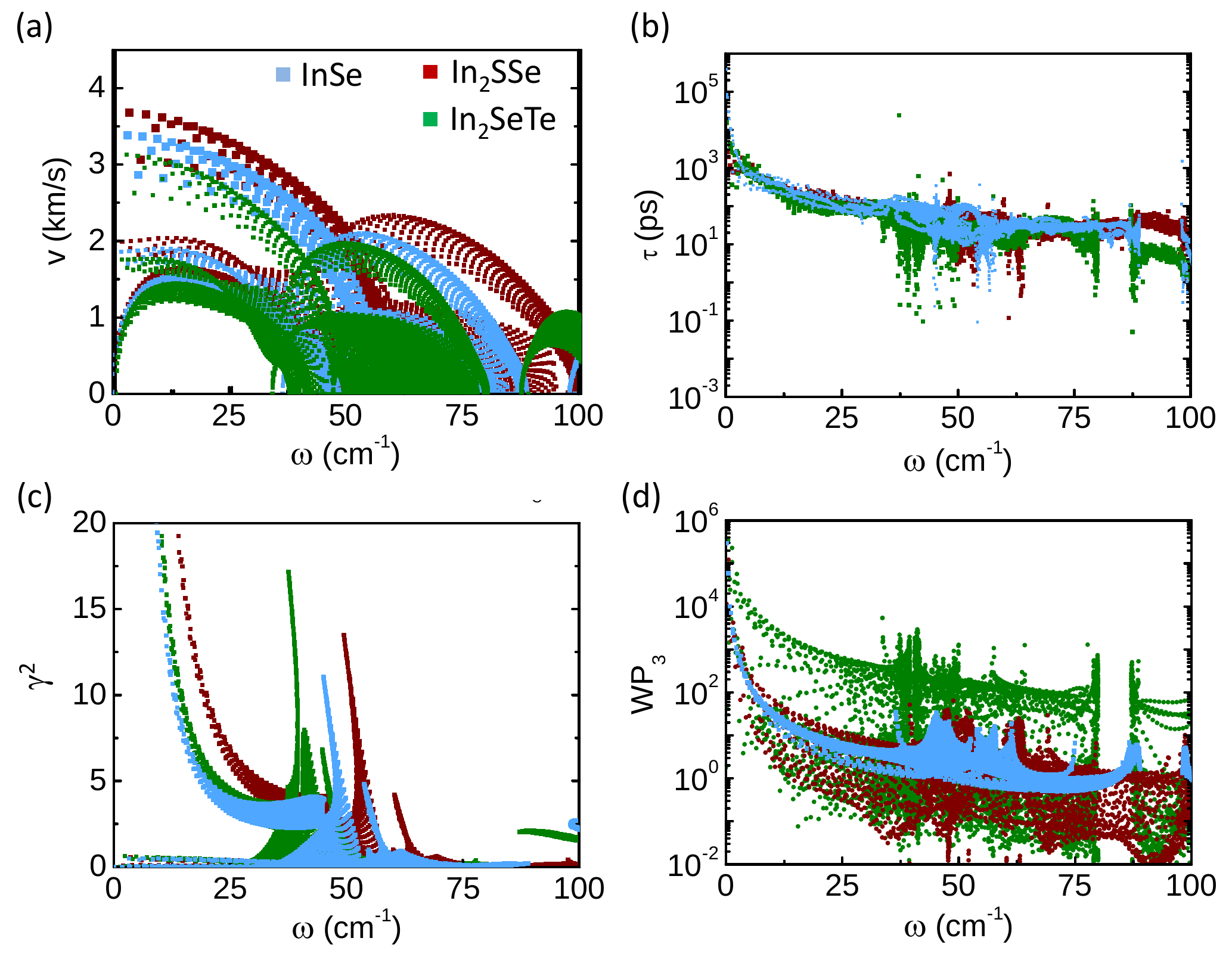}}
\caption{At room temperature, (a) mode group velocity $v_{\mathbf{q},s}$, (b) mode phonon life time $\tau_{\mathbf{q},s}$, (c) The square of mode gr\"{u}neisen parameter $\gamma^{2}_{\mathbf{q},s}$ and (d) phase space for anharmonic scattering $P_{3\mathbf{q},s}$ for monolayer InSe, In$_{2}$SSe and In$_{2}$SeTe.}
\label{wh4}
\end{figure}

To give a more deep understand of heat transport in monolayer InSe and its Janus structures, we analyzed every term in determining the $\kappa$ in Eq.~\ref{kappa}.
The mode group velocity $v_{\mathbf{q},s}$ in the frequency zone of $0\sim 100$ cm$^{-1}$ are displayed in Fig.~\ref{wh4}(a). The overall $v_{\mathbf{q},s}$ of monolayer In$_{2}$SSe and In$_{2}$SeTe is higher and lower than that of monolayer InSe, due to the smaller and larger atomic mass than that of InSe, respectively. Figure ~\ref{wh4}(b) shows the mode relaxation time $\tau_{\mathbf{q},s}$ at room temperature. The overall $\tau_{\mathbf{q},s}$ of monolayer InSe, In$_{2}$SSe and In$_{2}$SeTe have the same order of magnitude at low-frequency zone of $\omega<25$ cm$^{-1}$. However, the $\tau_{\mathbf{q},s}$ of Janus structures is smaller than that of monolayer InSe at the frequency between 25 and 75 cm$^{-1}$.
It was found that the anharmonic scattering ($1/\tau_{\mathbf{q},s}^{an}$) dominate the total phonon lifetime. In contrast, the isotopic impurities scattering, which is inversely proportional to atomic mass,~\cite{li2014shengbte} has a small influence on $\kappa$, e.g. the room-temperature $\kappa$ of monolayer In$_{2}$SSe will increase only by 1.26\% if we excluded the $1/\tau_{\mathbf{q},s}^{iso}$ from the total scattering rate $1/\tau_{\mathbf{q},s}$.

The anharmonic $1/\tau_{\mathbf{q},s}^{an}$ depends on the square of Gr\"{u}neisen parameter $\gamma_{\mathbf{q},s}^{2}$ and weighted phase space $W_{\mathbf{q},s}$.~\cite{li2014shengbte} The former and latter one represents the anharmonicity strength and the number of channels available for a phonon to get scattered, respectively.~\cite{weightspace,li2014shengbte}
The increase of $\gamma^{2}_{\mathbf{q},s}$ or $W_{\mathbf{q},s}$ has a decreased effect on $\tau_{\mathbf{q},s}$.

The overall $\gamma^{2}_{\mathbf{q},s}$ of monolayer InSe is smaller than that of its Janus derivatives (see Fig.~\ref{wh4}(c)). The breaking of mirror symmetry in Janus structures leads to the asymmetric chemical bonding and charge density distribution along the out-of-plane direction, which will increase the bond anharmonicity.~\cite{C7CP02486J,bonding}
Figure~\ref{wh4}(d)) shows that the sequential of $W_{\mathbf{q},s}$ from high to low is $W$(In$_{2}$SeTe)$>$$W$(InSe)$>$$W$(In$_{2}$SSe). $W_{\mathbf{q},s}$ of monolayer InSe, which has the mirror symmetry, is between that of monolayer In$_{2}$SSe and In$_{2}$SeTe. The phase space for anharmonic scattering decreases as the overall phonon frequency scale increases (see Fig.~\ref{wh1}(e)), consistent with the observation in many bulk materials.~\cite{Lindsay2008} As a result, $\tau_{\mathbf{q},s}$ of monolayer InSe is larger than that of
monolayer In$_{2}$SSe at frequency between 50 and 65 cm$^{-1}$ (see Fig.~\ref{wh4}(b) and Fig. S3), due to the
integral effect of $\gamma^{2}_{\mathbf{q},s}$, and $W_{\mathbf{q},s}$. In the end, monolayer In$_{2}$SSe, with smaller atomic mass, larger $v_{\mathbf{q},s}$ but a smaller $\tau_{\mathbf{q},s}$, has a slightly larger $\kappa$ than that of monolayer InSe. On the other side, the larger atomic mass, smaller $v_{\mathbf{q},s}$, larger $\gamma^{2}_{\mathbf{q},s}$ and $W_{\mathbf{q},s}$ leads to the smaller $\tau_{\mathbf{q},s}$ and $\kappa$ in monolayer In$_{2}$SeTe than InSe.

Besides the heat transport, we also calculated the carrier mobility $\mu$ based on the deformation potential theory (see computational details and Fig. S4 in SM).~\cite{deformation} The band structure indicates that electron effective mass is smaller than hole one which has Mexican-hat valence band edge (see Fig. S2), consistent with previous work.~\cite{piezoelectric} The parameters involved in electron mobility are shown in Table~\ref{table3}.
The order of $C_{\rm 2D}$ from large to small is $C_{\rm 2D}($In$_{2}$SSe$)>C_{\rm 2D}($InSe$)>C_{\rm 2D}($In$_{2}$SeTe), consistent with the order of $d_{ac}$ and $v_{\mathbf{q},s}$ (see Table~\ref{table1} and Fig.~\ref{wh4}(c)).

The electron effective mass $m^{*}$ of monolayer In$_{2}$SSe and In$_{2}$SeTe is larger and smaller than that of monolayer InSe, respectively. The orbital analysis indicates that the electronic states at conduction band minimum (CBM) of monolayer InSe is dominated by In-$5s$ orbital with symmetry spatial distribution (see Fig. S5). In contrast, the electronic states at CBM of monolayer In$_{2}$SSe and In$_{2}$SeTe is mainly composed of $5s$ orbital of In atom on the S and Se side, respectively. Taking monolayer In$_{2}$SeTe as an example, the Se side experiences aforementioned tensile strain compared to that of monolayer InSe.
Actually, if we applied a biaxial tensile strain of $2\%$ to the lattice of monolayer InSe, the $d_{In-Se}$ becomes close to that of monolayer In$_{2}$SeTe. Meanwhile, the $m^{*}$ changes from 0.188 $m_{0}$ to 0.174 $m_{0}$, consistent with the $m^{*}$ of In$_{2}$SeTe (Table~\ref{table3}). Similarly, $d_{In-S}$ of monolayer In$_{2}$SSe is about 2\% larger than that of monolayer InS. The calculated $m^{*}$ of monolayer InS under 2\% tensile strain is 0.227 $m_{0}$, which can also explain the larger $m^{*}$ of monolayer In$_{2}$SSe than that of monolayer InSe.

The $\mu$ of monolayer In$_{2}$SSe and In$_{2}$SeTe is smaller and larger than that of monolayer InSe, which can be attributed to larger $m^{*}$ and smaller $E_{1}$, respectively. Moreover, monolayer In$_{2}$SeTe has a direct band gap of 1.8 eV and thereby is more favorable for optoelectronics application than monolayer InSe and In$_{2}$SSe which have indirect band gaps (see Fig. S2),

\begin{table}[tbp!]
\caption{\label{table3}
The electron effective mass $m^{*}(1/m_{0})$ along $x$ and $y$ axis, 2D elastic module $C_{\rm 2D}$ (J/m$^{2}$), deformation potential constant $E_{1}$ (eV) and room-temperature $\mu$ (cm$^{2}$/V/s) of monolayer InSe and its Janus derivatives along the $x$ axis. The previous result of Ref. 34 is listed for a comparison.}
\begin{ruledtabular}
\begin{tabular}{cccccccc}
Type    &$m^{*}_{x}/m_{0}$ &$m^{*}_{y}/m_{0}$ &$C_{\rm 2D_x}$ &$|E_{1x}|$ & $\mu_{x}$\\
\hline
InSe            &0.181  &0.182    &49.21        &5.815   &943.3 \\
                &[0.177]~\cite{Chang2019}   &[0.182]~\cite{Chang2019} & & & \\
In$_{2}$SSe	    &0.211	 &0.212	  &52.69          &5.331   &884.8 \\
In$_{2}$SeTe	&0.175	 &0.176	  &44.22          &5.075   &1190.6 \\
\end{tabular}
\end{ruledtabular}
\end{table}

Based on the first-principles calculations, we investigated the thermal and electronic transport properties of monolayer InSe, In$_{2}$SSe and In$_{2}$SeTe. One side of monolayer In$_{2}$SSe and In$_{2}$SeTe undergo tensile or compressive strain. A distinguishable $A_{1}$ peak in the Raman spectra of monolayer In$_{2}$SSe and In$_{2}$SeTe was identified due to the breaking of mirror symmetry. The room-temperature $\kappa$ of monolayer InSe, In$_{2}$SSe and In$_{2}$SeTe is 44.6, 46.9 and 29.8 W/(m$\cdot$K), respectively. Though with smaller atomic mass, the $\kappa$ of monolayer In$_{2}$SSe is comparable to that of monolayer InSe, due to the competition effect between phonon group velocity and lifetime. Moreover, the electron mobility of monolayer In$_{2}$SeTe is higher than that of monolayer InSe, due to that smaller electron effective mass and deformation potential, respectively. Our work facilitates the understanding of the heat and carrier transport properties of 2D Janus group-III chalcogenides and offers the theoretical support to the corresponding device designment in future.

\section*{Supplementary Material}
See Supplemental Material for the computational details; The convergence test of $\kappa$ with respect to the Q-grid size and the force cut-off; The band structure and the distribution of electronic states; The fitting parameters for the mobility.

\begin{acknowledgments}
The numerical calculations in this paper have been done on the supercomputing system in the High Performance Computing Center of Yanshan University. This work was supported by the the Specialized Research Fund for the Doctoral Program of Higher Education of China (Grant No.2018M631760), the Project of Heibei Educational Department, China (No. ZD2018015 and QN2018012).

\end{acknowledgments}

\nocite{*}
\bibliography{ThInSe}
\end{document}